\documentclass[11pt, oneside]{article}   	
\usepackage{geometry}                		
\usepackage{graphicx}				
\usepackage{amssymb}

\usepackage{newtxtext}
\usepackage{newtxmath}
\usepackage{natbib}
\usepackage{hyperref}
\usepackage{epstopdf, epsfig}
\usepackage[utf8]{inputenc}
\usepackage{amsmath}
\usepackage{color}
\usepackage{listings}
\usepackage{float}
\usepackage{xcolor}
\hypersetup{
    colorlinks = false,
    urlcolor   = blue,
    citecolor  = black,
}

\newcommand{\RomanNumeralCaps}[1]
\linenumbers

\title{\textbf{A physical model for indirect noise in non-isentropic nozzles: Transfer functions and stability}}

\author{Animesh Jain$^1$ \& Luca Magri$^{2,1}$\footnote{l.magri@imperial.ac.uk} \\ 
 \small{$^1$ Department of Engineering, University of Cambridge,
Cambridge CB2 1PZ, UK} \\
\small{$^2$ Imperial College London, Aeronautics Department, London, UK}}

\date{}

\begin{document}
\maketitle

\begin{abstract}
We propose a mathematical model from physical principles to predict the sound generated in nozzles with dissipation.
The focus is on the sound generated from the acceleration of temperature inhomogeneities (also known as entropy waves), which is referred to as {\it indirect noise}. 
First, we model the dissipation caused by flow recirculation and wall friction with a friction factor, which enables us to derive quasi-one-dimensional equations {from conservation laws}. {The model is valid for both compact nozzles and nozzles with a  spatial extent.}  
{Second,} the predictions from the proposed model are compared against the experimental data available in the literature. 
{Third,} we compute the nozzle transfer functions for a range of Helmholtz numbers and friction factors. 
It is found that the friction and the Helmholtz number have a significant effect on the gain/phase of the reflected and transmitted waves. 
 {The analysis is performed from subsonic to supersonic regimes (with and without shock waves). The acoustic transfer functions vary significantly because of non-isentropic effects and the Helmholtz number, in particular, in the subsonic-choked regime. } 
Finally, we calculate the effect that the friction of a nozzle guide vane has on thermoacoustic stability. 
It is found that the friction and the Helmholtz number can change thermoacoustic stability from a linearly stable regime to a linearly unstable regime. 
The study opens up new possibilities for the accurate prediction of indirect noise in realistic nozzles with implications on both noise emissions and thermoacoustic stability. 
\end{abstract}

\section{Introduction}
In order to reduce the harmful effects of noise pollution generated by aircraft engines, manufacturers are striving to make engines less noisy, whilst keeping pollutant emissions minimal~\citep{correa1998power,hansell2013aircraft}. 
Whereas there has been a significant reduction in fan and jet noise, combustion noise, which is generated in the gas turbine combustor, has become a significant noise source in aircraft with low-emission engines~\citep{dowling2015combustion}. 
 %
 
The unsteady combustion process in an engine combustor is a source of both direct and indirect noise. 
On the one hand, the sound generated by the unsteady heat released by the flame, which leads to a volumetric contraction and expansion of the gas, is referred to as direct combustion noise~\citep[e.g.,][]{strahle1976noise}. 
On the other hand, the sound generated by the acceleration of flow inhomogeneities through the nozzle vane is referred to as indirect noise \citep[e.g.,][]{marble1977acoustic, cumpsty1979jet,polifke2001constructive, sattelmayer2003influence, goh2013influence, duran2013solution, motheau2014mixed, morgans2016entropy, magri2016compositional}.   
Depending on the flow inhomogeneity, indirect noise is further categorized as
(i)  entropy noise, when it is caused by temperature inhomogeneities \citep[e.g.,][]{marble1977acoustic}; 
(ii) compositional noise, when it is caused by compositional inhomogeneities~\citep[e.g.,][]{magri2016compositional, magri2017indirect}; 
and 
(iii) vorticity noise, when it is caused by velocity gradients~ \citep[e.g.,][]{howe1975contributions}. 
In low Mach numbers, vorticity noise is typically negligible~\citep{dowling2015combustion}. 
In contrast to direct noise, a complete understanding of indirect noise is yet to be  achieved \citep{ihme2017combustion,haghiri2018sound,tam2019combustion}. 
%
With a focus on entropy noise, a low-order model was proposed by~\citet{marble1977acoustic} for a system in which the nozzle length is negligible as compared to the wavelength of impinging disturbances (compact nozzle). 
Under the assumption of isentropic flow, the acoustic transfer functions were obtained by formulating jump conditions that conserve mass, total temperature and entropy~\citep{marble1977acoustic,cumpsty1977interaction}. 
%
When the nozzle spatial extent is not negligible with respect to the acoustic wavelength (non-compact nozzle), a shift in phase between inlet and outlet waves arises~\citep{marble1977acoustic}. 
The importance of the non-compact assumption was investigated and modelled by  \citet{leyko2009comparison}, among others.   
\citet{duran2013solution}, who obtained the acoustic and entropic transfer functions semi-analytically, showed that entropy noise  decreases with the Helmholtz number in choked nozzles. 
In addition to contributing to noise emissions, entropy noise can affect the combustor's thermoacoustic stability. 
When the acoustic waves that reflect off the nozzle guide vane are sufficiently in phase with the heat released by the flame in the combustor, a self-sustained thermoacoustic oscillation can arise~\citep{polifke2001constructive, goh2013influence, motheau2014mixed}.
Thermoacoustic instabilities are unwanted phenomena in gas turbines because they can lead to structural damage, which can reduce the combustor's lifetime and operability~\citep{dowling2015combustion}. 
%

The afore-mentioned studies in indirect-noise emissions and thermoacoustic stability assumed the nozzle flow to experience no losses in the stagnation pressure (isentropic assumption). 
However, in real situations, the flow is non-isentropic because of losses due to viscosity and recirculation zones~\citep{lieuwen2012unsteady}. 
In direct noise, non-isentropic effects were modelled in orifice plates~\citep{durrieu2001quasisteady}, and subsonic nozzles terminating in a duct \citep{howe1979attenuation, bechert1980sound,cummings1983high}. 
These studies were extended to a high-frequency regime to validate the acoustic transfer functions with experimental data \citep[e.g.,][]{dowling1992sound, durrieu2001quasisteady, bellucci2004numerical, yang2016analytical}. 
Recently,~\citet{de2017detection,de2017measurements}  showed a substantial mismatch between experimental and analytical acoustic/entropic transfer functions, when the latter are computed under the isentropic assumption in subsonic conditions. 
Indeed, the indirect noise emitted by the nozzle is markedly underestimated, whereas the indirect sound reflected off the nozzle is overestimated~\citep{de2019generalised}. 
 {With an elegant heuristic argument} and by introducing a semi-empirical parameter, i.e., the equivalent orifice area, \citet{de2019generalised} proposed a non-isentropic model to predict entropic-acoustic transfer functions for subsonic to sonic throat conditions in a compact nozzle. 
With an {\it ad-hoc} calibration of the equivalent orifice area, the model predictions compared favourably with experimental data. 
The impact of the isentropic assumption was found to be crucial. 
 {The model proposed was limited to (i) compact nozzles (and hence, low frequencies);  (ii) subsonic regimes; and (iii) heuristic arguments, i.e., the semi-empirical parameter assumed a portion of the nozzle to be replaced with an orifice plate. 
In this work, we derive the model from conservation laws and generalize it to non-compact nozzles and supersonic regimes.}

The objective is of this paper is threefold. 
We 
(i) propose a physical model from conservation laws, which captures the main non-isentropic source through friction, and generalizes to non-compact nozzles; 
 {(ii)  analyse the effect of non-isentropicity on the acoustic and entropic transfer functions for subsonic and supersonic regimes (with and without shock);}
and 
(iii) show that thermoacoustic stability is affected by non-isentropic effects in the nozzle.
For this, a convergent-divergent nozzle is numerically investigated. 
The results are compared with existing experimental data~\citep{de2019generalised}.
The paper is structured as follows. 
Section~\ref{sec:MathModel} introduces the mathematical model with physical interpretation of the equations. 
Section~\ref{subsec:betavsf} physically interprets the semi-empirical parameter. 
Section~\ref{sec:results} shows the acoustic and entropic transfer functions in a subsonic regime. 
Section~\ref{sec:results_sup} shows the indirect-noise transfer functions in a  supersonic regime. 
Section~\ref{sec:tas} shows the thermoacoustic stability for different levels of nozzle non-isentropicity.  
Conclusions end the paper.


\section{Mathematical model}\label{sec:MathModel}
We consider a flow that is single component, ideal, and calorically perfect.
The effect of body forces and diffusion are assumed to be negligible. 
The flow evolves in an adiabatic nozzle, whose cut-off frequency is sufficiently high for the acoustics to be one dimensional.
The conservation of mass, momentum, and energy are, respectively~\citep[e.g.,][]{shapiro1954compressible} 
\begin{align}
    \frac{D\rho}{Dt} + \rho\frac{\partial u}{\partial x} + \frac{\rho u}{A}\frac{dA}{dx} &= {\Dot{S}_m}, \label{eq:dmdt_1}\\
    \frac{Du}{Dt} + \frac{1}{\rho}\frac{\partial p}{\partial x} &= {\Dot{S}_M}, \label{eq:dmdt_132412}\\
    T\frac{Ds}{Dt} &= {\Dot{S}_s},\label{eq:dsdt_1}
\end{align}
where 
$x$ is the longitudinal coordinate of the nozzle, 
$\rho$ is the density, 
$u$ is the velocity, 
$p$ is the pressure, 
$T$ is the temperature, 
$s$ is the entropy, 
and $A$ is the cross-sectional area of the duct. 
The right-hand side terms, ${\Dot{S_j}}$, are the source terms of mass, momentum, and entropy, respectively. 
The thermodynamic variables satisfy Gibbs' equation 
\begin{align}\label{eq:ge1}   
{ds} & = c_p\frac{dT}{T} - R\frac{dp}{p},
\end{align}
where $R$ is the gas constant,
and $c_p$ is the specific heat at constant pressure. 
We assume that the mass source is  zero, $\Dot{S}_m = 0$, in \eqref{eq:dmdt_1}.

In the energy equation, the dissipation is modelled as an entropy source. 
Introducing the compressibility factor 
\begin{align}
\Lambda\equiv {1 + \frac{\gamma - 1}{2}M^2}, 
\end{align} 
and substituting the thermodynamic relations $T = {T_0}/\Lambda$ and $p = {p_0}/{\Lambda}^\frac{\gamma}{\gamma - 1}$ in \eqref{eq:ge1}, where
 $\gamma$ is the heat-capacity ratio, yield 
\begin{align}\label{eq:ge2}  
    ds & = c_p \frac{dT_0}{T_0} - R\frac{dp_0}{p_0}, 
\end{align}
which, under the adiabatic assumption, $dT_0 = 0$, simplifies to  
\begin{align}\label{eq:EST1}
    \Dot{S}_s &= -RT\frac{D\log p_0}{Dt}.
\end{align}
Physically, a reduction in the stagnation pressure increases the entropy production proportionally 
to the temperature of the flow. 
If the flow is isentropic, the stagnation pressure is constant, and vice versa.
%
%
In the momentum equation, the dissipation is modelled as a momentum source   
\begin{align}
    \Dot{S}_M = - \frac{4f}{D}\frac{u^2}{2}, 
\end{align}
where  $f = \tau_w / ({\rho u^2}/{2})$ is the Fanning friction factor, in which $\tau_w$ is the shear stress~\citep{shapiro1954compressible}.  
The friction decreases the stagnation pressure as
\begin{align}\label{eq:fandpop1}
    {d\log p_0} = -\frac{\gamma M^2}{2} \frac{4f}{D} dx,
\end{align}
where $D$ is the nozzle diameter.
 {In quasi-one-dimensional models, the flow variables at $x$ are averages across the cross-section $A(x)$. This means that the friction factor can  account for wall friction and two (or three) dimensional effects, such as flow recirculation, in a cross-averaged sense.} (These effects can be accurately captured in ducts with a slowing-varying cross-sectional area, for which the quasi-one dimensional assumption holds.) 
The Mach number, $M$, is governed by~\citep{shapiro1953dynamics} 
%
\begin{align}\label{eq:dm2bm21_1}
    \frac{dM^2}{M^2} = \frac{\Lambda}{1 - M^2}\left[-2\frac{dA}{A} + \gamma M^2  \frac{4f}{D}{dx}\right]. 
\end{align}
Equation~\eqref{eq:dm2bm21_1} is the extension of the Fanno flow to variable-area ducts. 
\begin{figure}
\centering    
\includegraphics[width=0.8\textwidth]{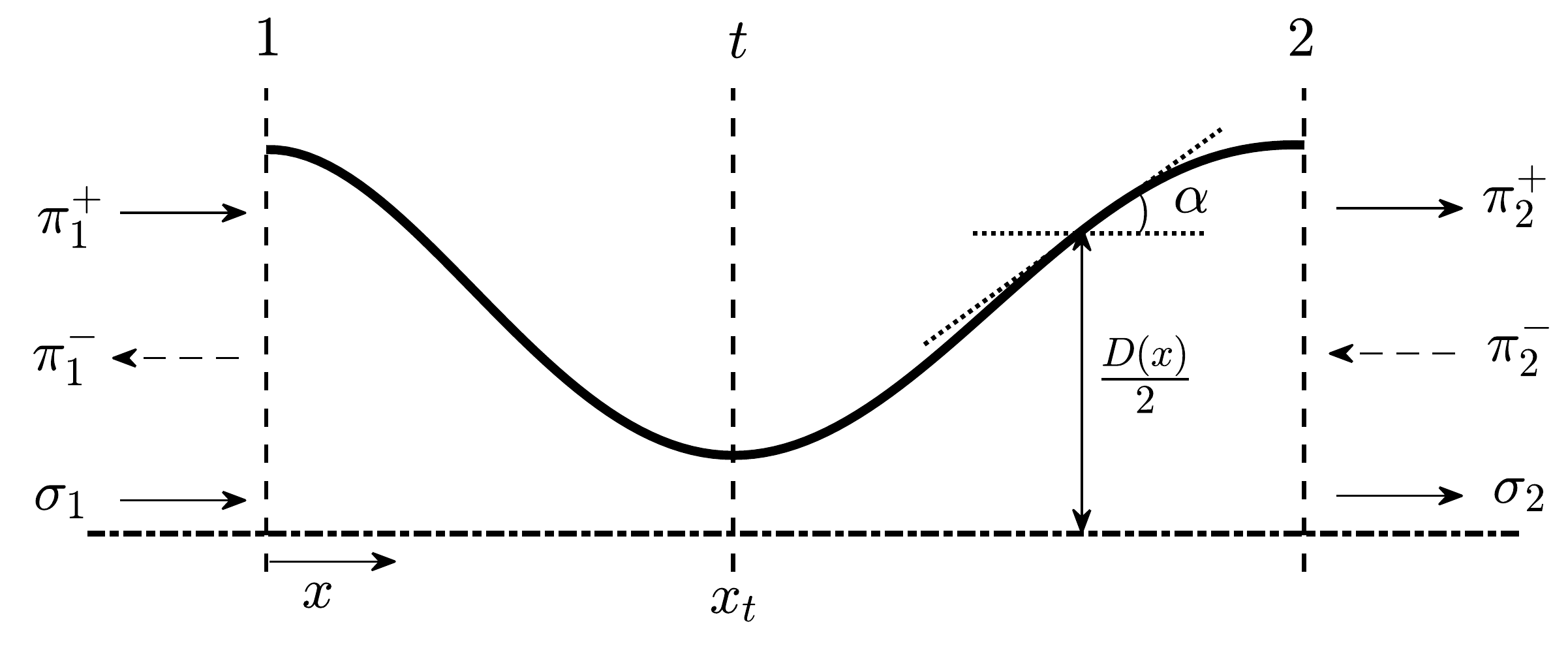}
\caption{Schematic of a generic  nozzle with nomenclature.}
\label{fig:nozz_cd}
\end{figure}
%
 {We start with an arbitrary nozzle geometry as shown in Figure \ref{fig:nozz_cd}. Using the nozzle profile, at any location $x$, $4dx/D = dA/(A \tan\alpha)$, where $D$ is the diameter of the cross section and $\tan\alpha$ is the spatial derivative of the nozzle profile at that location}, 
which reduces~\eqref{eq:dm2bm21} to
\begin{align}\label{eq:dm2bm21}
    \frac{dM^2}{M^2} = \left(\frac{\Lambda}{1-M^2}\right)\frac{\zeta}{\tan\alpha}\frac{dA}{A}
\end{align}
where, for brevity, we define the competition factor $\zeta \equiv  f\gamma M^2 - 2 \tan\alpha$, which quantifies the competition between the dynamics dictated by mass conservation~\eqref{eq:dmdt_1}, through the nozzle geometry, $\alpha$, and  
 the dynamics dictated by momentum conservation~\eqref{eq:dmdt_132412}, through the friction, $f$. 
  If the two mechanisms are in balance, $\zeta=0$, the flow evolves at a constant Mach number. 
The stagnation pressure is
\begin{align}\label{eq:dpobpomsq}
    \frac{D\log p_0}{Dt} = -\frac{f}{\zeta}\left(\frac{\gamma(1 - M^2)}{2\Lambda}\right)  \frac{DM^2}{Dt}. 
\end{align}
(The condition $\zeta=0$ is not a singularity because \eqref{eq:dpobpomsq} $\rightarrow$ \eqref{eq:fandpop1} as $\zeta\rightarrow 0$.)
To physically interpret~\eqref{eq:dpobpomsq}, we consider a symmetric  {linear geometry} nozzle in which $\alpha_c=-\alpha_d$, where $c$ and $d$ denote the convergent and divergent sections, respectively. 
As defined in Figure~\ref{fig:nozz_cd}, $\alpha_c<0$. 
From~\eqref{eq:dm2bm21} and \eqref{eq:dpobpomsq}, 
changes in the stagnation pressure for the same Mach number, $M$, are related by 
\begin{align} \label{eq:fhwfqh3}
\frac{(d\log p_0)_d}{(d\log p_0)_c} \sim \frac{\zeta_c}{\zeta_d} > 1. 
\end{align}
This means that the flow experiences greater pressure stagnation losses in the divergent section of the nozzle for the same Mach number, 
which is physically consistent with experiments~\citep{de2019generalised}. 
Finally, the entropy source term~\eqref{eq:EST1} can be expressed as a function of the friction factor by substituting~\eqref{eq:dpobpomsq} into~\eqref{eq:EST1}.  
%
In the limit of constant-area ducts, $\alpha\rightarrow0$, the equations tend to the Fanno flow.
In the limit of frictionless nozzles, $f \rightarrow 0$, the equations tend to the isentropic model. 
\subsection{Linearization}
We model the acoustics as linear perturbations to a mean flow. 
For this, we decompose a generic flow variable, $v$, as $v\rightarrow {v}(x) + v^{\prime}(x,t)$, where ${v}(x)$ is the steady mean flow component, and $v^{\prime}(x,t)$ is the first-order perturbation. 

Assuming a constant friction factor, $f$, the analytical integration of~\eqref{eq:dm2bm21} and \eqref{eq:fandpop1}, provides the nonlinear algebraic equations for the mean-flow Mach number, stagnation pressure and entropy, respectively 

\begin{align}
    \label{eq:a2ba1m2bm1}
    \frac{A_2}{A_1} &= \frac{M_1}{M_2} \left(\frac{\Lambda_2}{\Lambda_1}\right)^\frac{(\gamma + 1) \tan \alpha}{2\kappa} \left(\frac{\zeta_1}{\zeta_2}\right)^\frac{f\gamma - 2 \tan\alpha}{2\kappa}, \\ 
  \frac{p_{02}}{p_{01}} &= \left(\frac{\Lambda_2}{\Lambda_1}\right)^\frac{f\gamma(\gamma + 1)}{2(\gamma - 1)\kappa} \left(\frac{\zeta_1}{\zeta_2}\right)^\frac{f\gamma - 2 \tan\alpha}{2\kappa}, \\
  \label{eq:delsbcpwithf}
  \frac{\Delta s}{c_p} & = \log\left(\frac{\Lambda_1}{\Lambda_2}\right)^\frac{f(\gamma + 1)}{2\kappa} \left(\frac{\zeta_2}{\zeta_1}\right)^{\frac{\gamma - 1}{\gamma}\frac{f\gamma - 2 \tan\alpha}{2\kappa}}. 
  %
\end{align}

where, $\kappa = f\gamma + (\gamma - 1)\tan\alpha$. 
(Solutions~\eqref{eq:a2ba1m2bm1}-\eqref{eq:delsbcpwithf} can be used for curved nozzle geometries by discretizing the nozzle shape piecewise linearly.) 
After some algebra, it can be shown that the linear perturbations are governed by
\begin{align}
  &\frac{  D}{D\tau}\left(\frac{p^{\prime}}{\gamma   p}\right) + \tilde{u}\frac{\partial}{\partial \eta}\left(\frac{u^{\prime}}{  u}\right) - \frac{  D}{D\tau}\left(\frac{s^{\prime}}{  c_p}\right) = 0, \label{eq:co2_1}\\
  &\frac{  D}{D\tau}\left(\frac{u^{\prime}}{  u}\right) + \frac{\tilde{u}}{  M^2}\frac{\partial}{\partial \eta}\frac{p^{\prime}}{\gamma  p} +  \left(2\frac{u^{\prime}}{  u} + (1 - \gamma)\frac{p^{\prime}}{\gamma   p} - \frac{s^{\prime}}{  c_p}\right)\left(\frac{\partial\tilde{u}}{\partial \eta} + \frac{4f}{D}\frac{\tilde{u}}{2} \right) = 0, \label{eq:mo2} \\ 
    %
    &\frac{D}{D\tau}\left(\frac{s^\prime}{ c_p}\right) = g(f,f^2,f^3, {o}(f^3)), \label{eq:dscpdt_1}
\end{align}
where the linearized entropy source term, $g(\cdot)$, is provided in Appendix~\ref{app:fihjrf329f3}, and $o$ is the little-O Landau symbol.
%
The linearized Gibbs' equation is~\citep{marble1977acoustic}
\begin{align}\label{eq:rho_chpt3b}
    \frac{\rho^{\prime}}{\rho} = \frac{p^{\prime}}{\gamma  p} - \frac{s^{\prime}}{ c_p}.  
\end{align}
The variables are non-dimensionalised as 
$\tau = tf_a$, $\eta = x/L$, 
$  M =   u/   c$ 
and $\tilde{u} =   u / c_{ref}$, 
where 
$f_a$ is the frequency of the advected perturbations entering the nozzle, 
$L$ is the axial length of the nozzle, 
and $c_{ref}$ is the reference speed of sound. 
Consequently, the non-dimensionalized material derivative becomes 
${D}/{D\tau} = He {\partial}/{\partial\tau} + \Tilde{u}{\partial}/{\partial\eta}$,  
where the Helmholtz number is defined as $He = {f_a L}/{c_{ref}}$.
Physically, the Helmholtz number is the ratio between advected perturbations and acoustic wavelengths. 
If the length of nozzle is small as compared to the wavelength of the acoustic and the entropy perturbations, 
the nozzle is compact ($He = 0$). 
Equation \eqref{eq:dscpdt_1} is derived by linearising \eqref{eq:dsdt_1} and substituting \eqref{eq:co2_1}-\eqref{eq:mo2}. 

In the limit of isentropic flow, $f = 0$, \eqref{eq:co2_1}, \eqref{eq:mo2}, \eqref{eq:dscpdt_1} tend to the  linearized Euler equations of~\citet{duran2013solution}. 
On the one hand, in a frictionless flow, the sound is generated by the advected inhomogeneities through the acceleration ${\partial\tilde{u}}/{\partial \eta}$ in \eqref{eq:mo2}. 
On the other hand, in non-isentropic flows, the friction term, $4f\tilde{u}/2D$, is a negative acceleration, which competes with the advection of the mean flow. 
In \S~\ref{sec:riemaninvar}, we recast the equations in Riemann invariants, which are necessary for the numerical procedure proposed. 

\subsubsection{Riemann Invariants} \label{sec:riemaninvar}
%
First, \eqref{eq:co2_1}-\eqref{eq:dscpdt_1} are Fourier transformed with the decomposition $\mathbf{q}(\tau,\eta) = \hat{\mathbf{{q}}}({\eta})\exp(2\pi i \tau)$.
Second, the primitive variables are decomposed in travelling waves, where 
the downstream (superscript $+$) and upstream (superscript $-$) propagating acoustic waves are 
    $\pi^\pm = 0.5\left[{p^\prime}/{(\gamma p)} \pm {u^\prime}/{ u} \right]$;  and the advected entropy wave is
   $ \sigma = {s^\prime}/{c_p}$. 
Third, the  equations are solved as a boundary value problem with boundary conditions specified for waves according to the transfer functions being evaluated.
This provides a system of four linear equations in the gradients of the four primitive variables, which is solved by inversion. 
The gradients of the primitive variables provide the gradients of the Riemann invariants at each axial location, which, in turn, are used to update the values of Riemann invariants.

The process is repeated until the boundary conditions are matched (Figure~\ref{fig:nozz_cd}).  
%
%
As a measure of direct noise, we use the acoustic-acoustic reflection coefficient, $\pi_1^- / \pi_1^+$ and  the acoustic-acoustic transmission coefficient, $\pi_2^+ / \pi_1^+$. 
As a measure of indirect noise, we use the entropic-acoustic reflection coefficient, $S_R={\pi_1^-}/{\sigma_1}$, and the entropic-acoustic transmission coefficient, $S_T={\pi_2^+}/{\sigma_1}$.

\section{Physical interpretation of semi-empirical non-isentropic models}\label{subsec:betavsf}
\begin{figure}
\centering    
\includegraphics[width=0.75\textwidth]{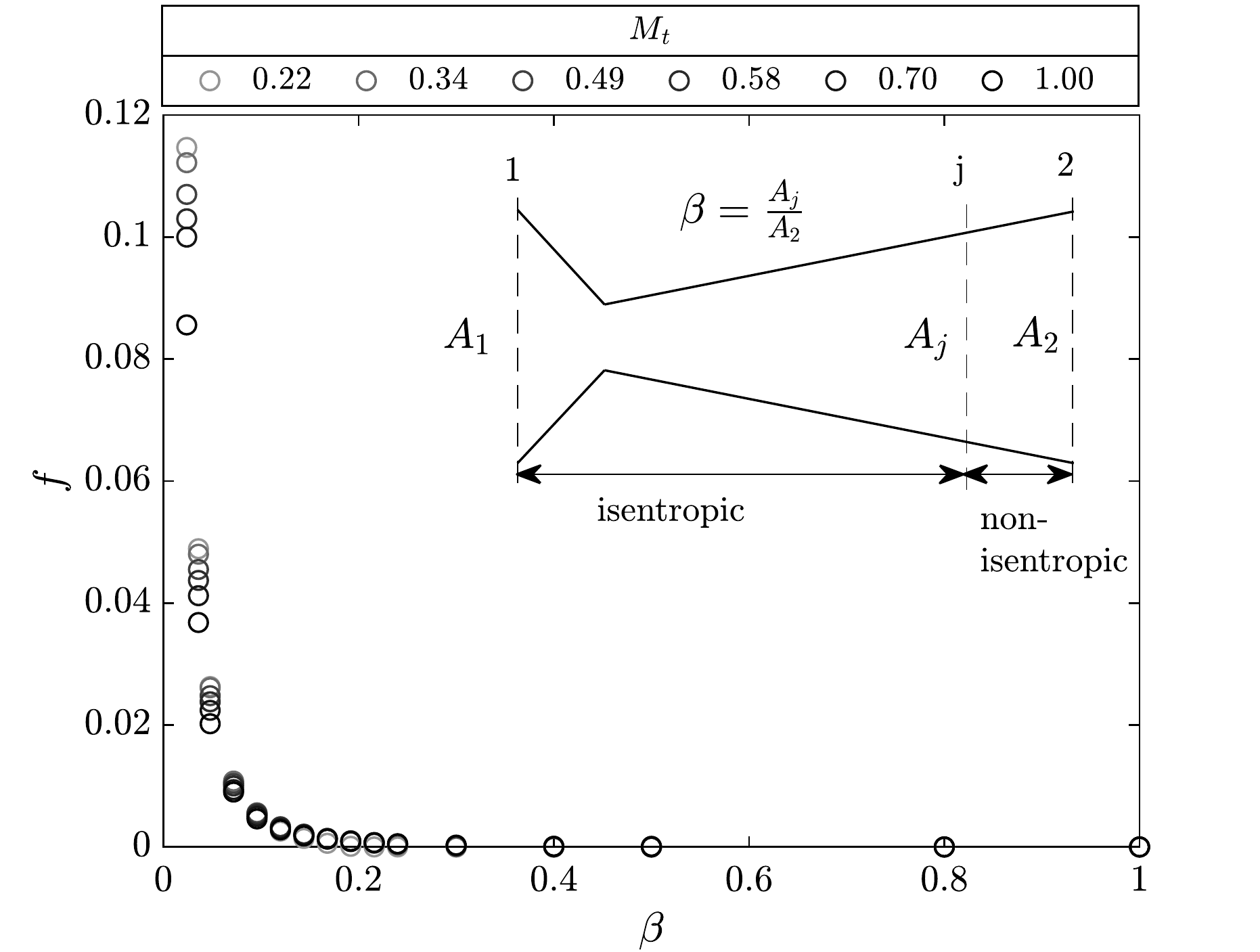}
\caption{Relation between the semi-empirical parameter, $\beta$, of the literature~\citep{de2019generalised} (inset), and physical friction, $f$, proposed in this paper.}
\label{fig:betavsf}
\end{figure}

In the literature, non-isentropic models for indirect noise are semi-empirical and valid for compact nozzles~\citep{de2019generalised}. 
The dissipation in the divergent section, which is caused by flow separation due to adverse pressure gradients, was modelled as an   orifice plate.
To do this, the non-isentropicity was embedded in one semi-empirical parameter, $\beta = {A_j}/{A_2}$, where 
 $j$ denotes the location at which non-isentropicity is assumed to begin (equivalent orifice area), 
 and $2$ denotes the nozzle exit (inset in Figure~\ref{fig:betavsf}). 
The non-isentropicity parameter, $\beta$, was assumed to be related to the loss in stagnation pressure using the Borda-Carnot equation, which describes losses in mechanical energy due to sudden flow expansion. 
As a function of the pressure loss coefficient,  \citet{de2019generalised} showed that $C_{p0} = {(p_{0,j} - p_{0,2})}/(0.5\rho_j u_j^2)=(1 - \beta)^2$, 
where $p_{0,2}$ is the stagnation pressure at the outlet, 
and 
$p_{0,j} = p_{0,1}$. 
The semi-empirical parameter, $\beta$, was calibrated experimentally by measuring the pressure loss as a function of the mass flow rate~\citep{de2019generalised}.

We propose a physical interpretation of the semi-empirical parameter, $\beta$, and a physical equation for it. 
Consistently with~\citet{de2019generalised}, we assume the nozzle to be divided into an isentropic part ($1 - j$), 
and a non-isentropic part ($j - 2$) ({Figure~\ref{fig:betavsf}}). 
Integrating Gibbs' equation and the energy equation in $j - 2$  
\begin{align}
    \int_{j}^{2}\frac{ds}{R} = \int_{j}^{2}\frac{(1 - M^2)}{\Lambda} \frac{d M}{M}  +  \int_{j}^{2}\frac{d A}{A}, 
\end{align}
yields 
\begin{align}\label{eq:b11}
    \beta = \left[\frac{M_2}{M_j}\left({\frac{\Lambda_j}{\Lambda_2}}\right)^{\frac{\gamma + 1}{2(\gamma - 1)}}\right] \exp{\left(-\frac{\Delta s}{R}\right)}. 
\end{align}
%
From~\eqref{eq:ge2}, the net change of entropy is ${\Delta s}/{R} = - \log ({p_{0,2}}/{p_{0,j}})$, which, in terms of the pressure coefficient, is
\begin{align}\label{eq:dsor}
    \frac{\Delta s}{R} = - \log\Bigg[1 - \left(\frac{C_{p0}\frac{1}{2}\gamma M_j^2 }{\Lambda_j}\right) \Bigg]. 
\end{align}
The properties at section $j$ are isentropic. 
For a fully isentropic nozzle $C_{p0} = 0$ ($\Delta s=0$), therefore,  section $j$ coincides with $2$ with $\beta$ becoming unity.  

To connect the semi-empirical non-isentropicity factor, $\beta$, with the  physical friction parameter, $f$,  from \eqref{eq:delsbcpwithf} and \eqref{eq:b11}, it can be shown that 
\begin{align} \label{eq:fi3rjhfiqjrf3}
    \beta = \frac{M_2}{M_j}\left({\frac{\Lambda_j}{\Lambda_2}}\right)^{\frac{\gamma + 1}{2(\gamma - 1)}} \left(\frac{\Lambda_1}{\Lambda_2}\right)^\frac{-f\gamma(\gamma + 1)}{2(\gamma - 1)\kappa} \left(\frac{\zeta_1}{\zeta_2}\right)^\frac{f\gamma - 2 \tan\alpha}{2\kappa}. 
\end{align}

As shown in Figure~\ref{fig:betavsf}, larger values of $\beta$ correspond to lower non-isentropicity, or, equivalently, lower values of the friction factor.  
For the nozzle geometry of~\citet{de2019generalised}, which is used in this study, 
the lowest value is $\beta = 0.024$, which is attained when the throat area is the equivalent orifice plate area. 
As the effect of non-isentropicity decreases, i.e., $\beta$ increases, the friction factor, $f$, becomes negligible. 
From~\eqref{eq:fi3rjhfiqjrf3}, for an isentropic nozzle, $\beta = 1$, the friction factor, $f$, is zero.  
The proposed model captures the variation of the friction factor with the throat Mach number.  
Physically, a flow with a higher throat Mach number needs less friction to generate the same entropy of a flow with a smaller throat Mach number. It can be shown from \eqref{eq:ge1}, \eqref{eq:fandpop1}, \eqref{eq:dsor} that 
\begin{align}
\frac{d s}{dM^2} \sim   - \frac{df}{dM^2} \sim  (1-\beta)^2, 
\end{align}
which means that, for a given non-isentropicity parameter, $\beta$,   friction needs to decrease as the Mach number increases. 
Similarly, the sensitivity of the friction factor to the throat Mach number quadratically decreases with $\beta$, which becomes negligible for $\beta \gtrsim 0.1$ (Figure~\ref{fig:betavsf}).  {Equation \eqref{eq:fi3rjhfiqjrf3} shows that the friction factor models (in a cross-averaged sense) the flow recirculation dissipation captured by the orifice-place parameter, $\beta$. }


\section{Indirect-noise transfer functions  {in a subsonic-choked regime}}\label{sec:results}
We calculate the transmission and reflection coefficients in a linear geometry nozzle in a subsonic (up to choked) regime. 
Both compact nozzles and  non-compact nozzles are analysed.
The predictions on the compact nozzle are validated against  experimental data available in the literature~\citep{de2019generalised}. 
%
The ambient temperature is $293.15$K, the exit pressure is $10^5$Pa, and $\gamma=1.4$. 
The nozzle has inlet and outlet diameters of $46.2$mm, throat diameter of $6.6$mm,  length of converging section of $24$mm, and length of divergent section of $230$mm. The angles of the convergent and divergent sections are $\alpha_c = -40^{\circ}$ and $\alpha_d = 4^{\circ}$, respectively (Figure~\ref{fig:nozz_cd}). The vena contracta factor  is  $\Gamma = 0.89$, which is defined as the ratio between the cross section area of the stream and the throat area~\citep{de2019generalised}.  

\subsection{Mean flow}\label{sec:meanflow}
\begin{figure}
\centering    
\includegraphics[width=1.0\textwidth]{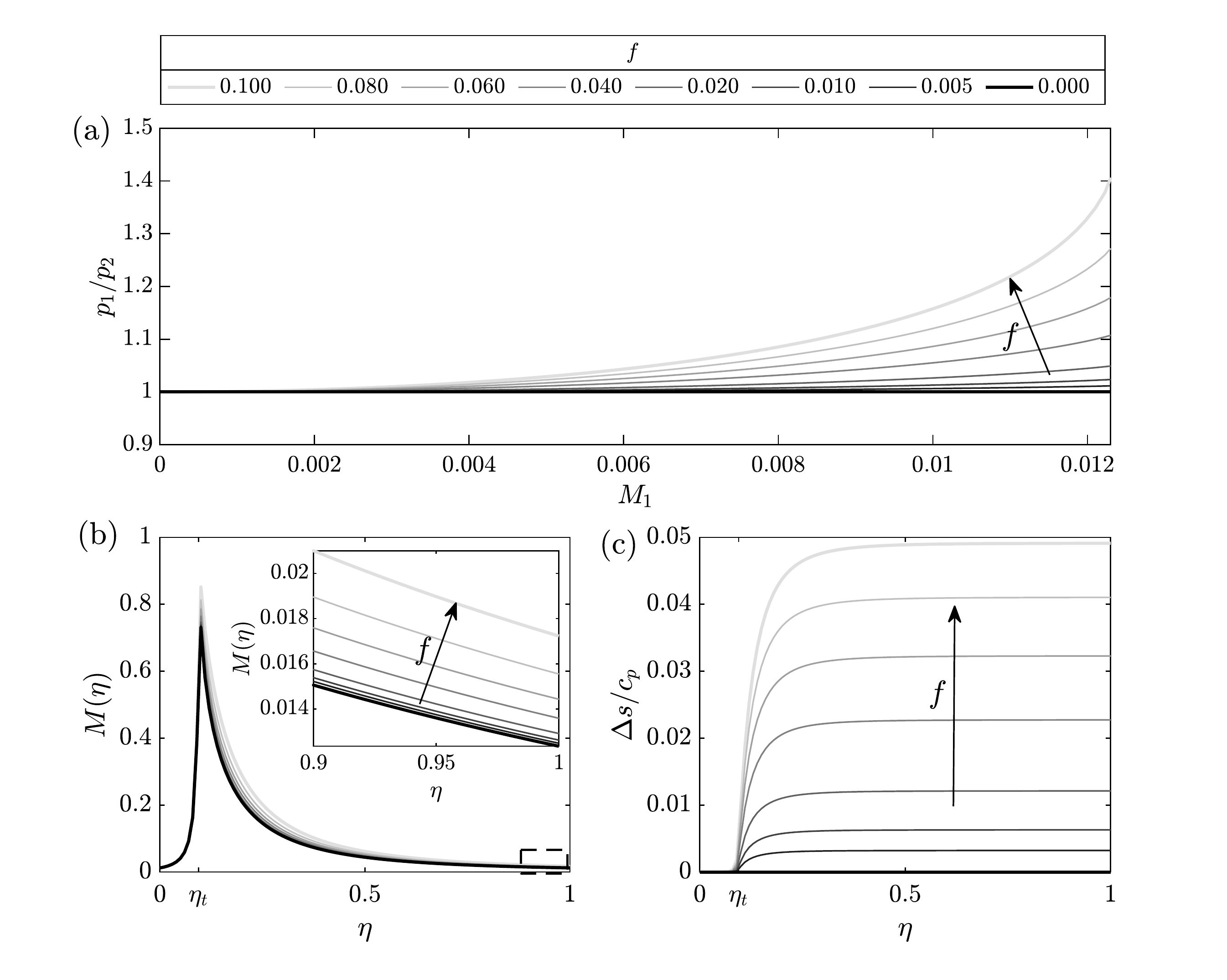}
\caption{Mean-flow. (a) Ratio of inlet and exit pressures, (b) Mach number and (c) entropy. 
The arrows indicate increasing friction, $f$. 
The normalised throat location is $\eta_t = 0.095$. } 
\label{fig:nozz_cd_1}
\end{figure}
Figure \ref{fig:nozz_cd_1} shows the mean-flow parametrized with the friction. 
As shown in panel (a), because the outlet pressure $p_2$ is fixed, to compensate for the pressure loss, the inlet pressure $p_1$ increases as the friction increases.
The effect becomes more evident as the inlet Mach number increases. 
As intuitively expected, higher friction results in larger pressure losses. 
%
As shown in panel (b),  as the friction increases, the outlet Mach number becomes larger than the inlet Mach number (inlet and outlet areas are equal). 
This is consistent with a Fanno flow-like behaviour.
As shown in panel (c), entropy is mostly generated near the throat, where the nozzle geometry switches from a converging (the flow is attached) to a diverging (the flow partly separates) regime. 
(Mathematically, the sign of the angle, $\alpha$, in \eqref{eq:delsbcpwithf} changes.) 
Physically, in the converging section, the only force opposing the flow is friction, but, in the divergent section, the force from the adverse pressure gradient also opposes the flow  motion, which makes it susceptible to separation. 
At the throat, the overall opposing force changes abruptly, which makes the losses large. 
This can also be seen from the stagnation pressure equation
\begin{align}\label{eq:stagpressuremach}
    \frac{dp_0}{p_0} = \frac{dp}{p} + \frac{\gamma M^2/2}{\Lambda}\frac{dM^2}{M^2}.
\end{align}
On the one hand, in an isentropic flow, the change in pressure term is in equilibrium with the inertia term (Mach number), thus, the stagnation pressure is constant. 
On the other hand, in a subsonic non-isentropic flow, the pressure gradient becomes adverse right after the throat ($dp>0$).  
This change of sign generates maximum entropy production (Figure \ref{fig:nozz_cd_1}(c)). 
\begin{figure}
\centering    
\includegraphics[width=1.0\textwidth]{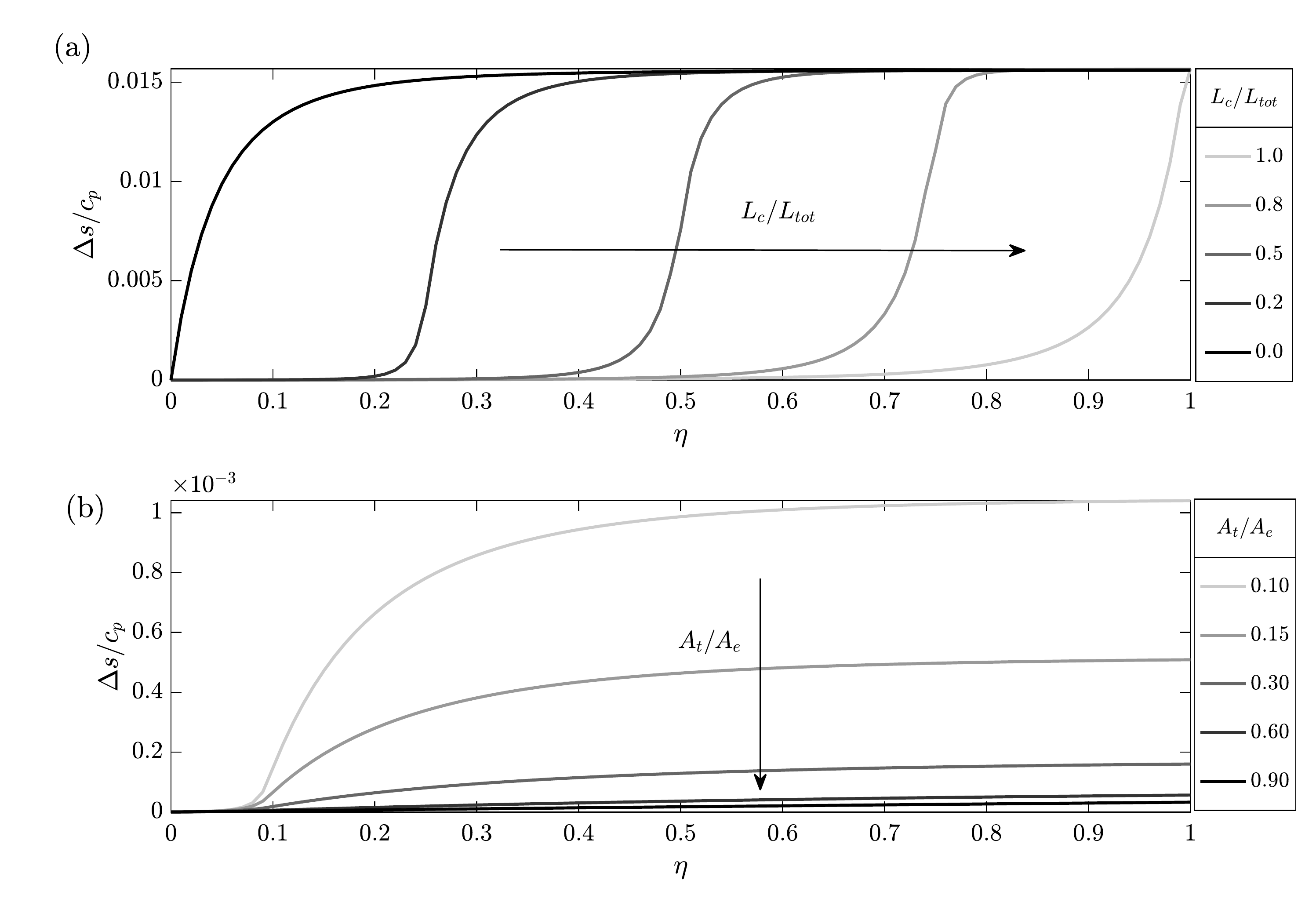}
\caption{Effect of geometry on the mean flow. (a) Converging-section-to-total-length ratio ($L_c/L_{tot}$) with $L_{tot}$ fixed.
(b) Throat-to-exit area ratio ($A_t/A_e$) with $A_e$ fixed. (In the geometry considered in this paper, the exit area is equal to the inlet area.) 
}
\label{fig:nozz_cd_2}
\end{figure}
Figure \ref{fig:nozz_cd_2}(a) shows the effect of of the nozzle geometry.  
As the location of the throat moves downstream, the highest entropy generation location also changes. 
The limiting curves correspond to a diffuser ($L_c/L_{tot} = 0$) and a converging nozzle ($L_c/L_{tot} = 1$), in which  
the maximum increase in entropy takes place at the inlet and outlet, respectively.  
Finally, as shown in Figure~\ref{fig:nozz_cd_2}(b), as the throat-to-exit-area ratio increases, the entropy production decreases because the flow undergoes a reduced separation.  
In conclusion, the location of the throat determines the location of maximum entropy generation, whereas the throat-to-exit area ratio determines the amount of entropy generated. 
Although quasi-one dimensional, the proposed model is able to capture the two dimensional dissipation effects in the divergent section, which are averaged across the section, through the friction factor. 
%
\subsection{Indirect noise in compact nozzles}
%
\begin{figure}
\centering    
\includegraphics[width=1.0\textwidth]{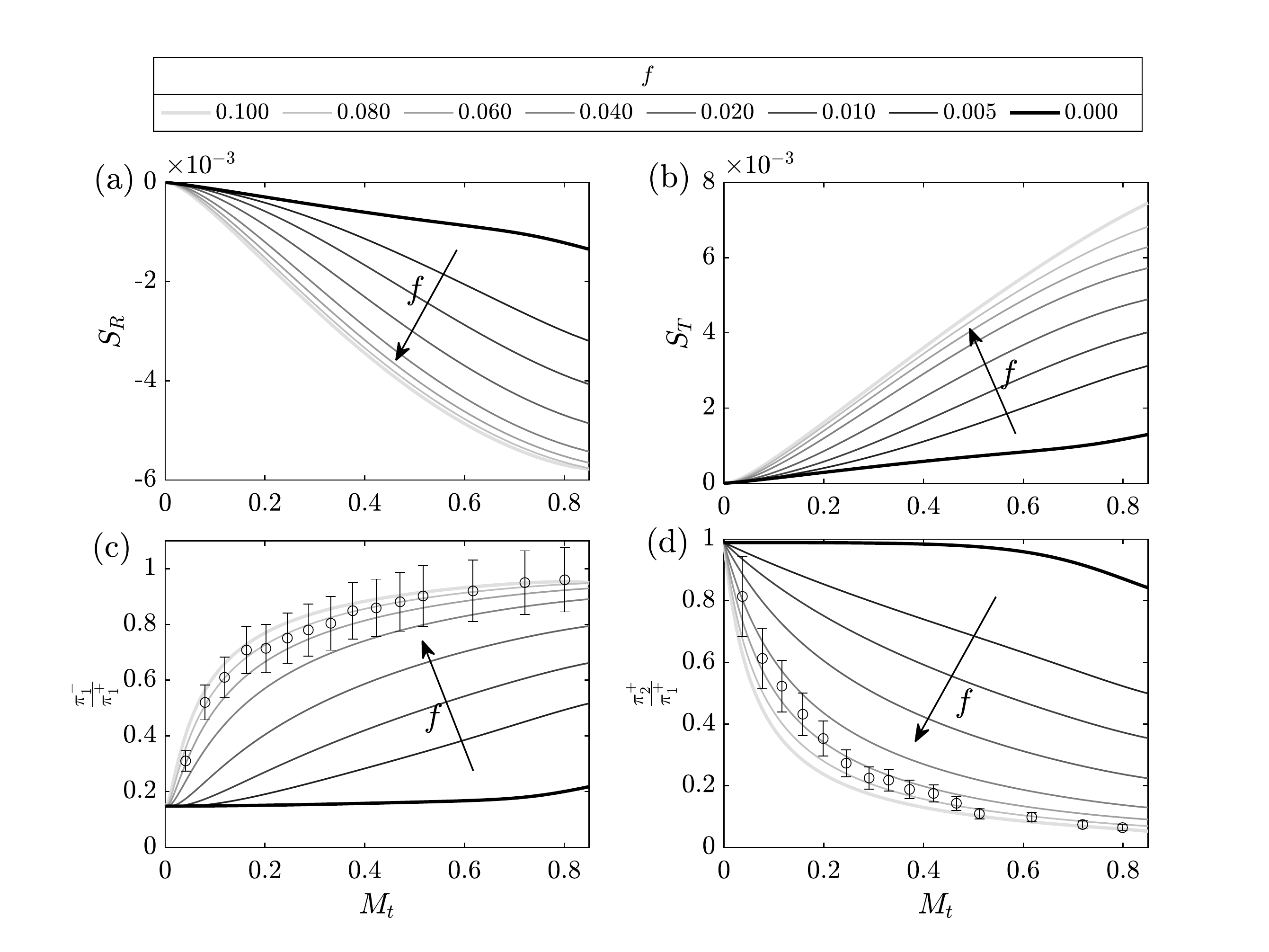}
\caption{Compact nozzle. Entropic-acoustic (a) reflection and (b) transmission coefficients for the (nearly) compact nozzle ($He = 0.0074$) of the experimental setup in~\citet{de2019generalised}.}
\label{fig:fhezero_1}
\end{figure}
Figure~\ref{fig:fhezero_1} shows (i) the entropic-acoustic reflection, $S_R$, and transmission, $S_T$, coefficients, and (ii) the acoustic-acoustic reflection, $\pi_1^-/\pi_1^+$, and transmission, $\pi_1^-/\pi_1^+$, coefficients in a nearly compact nozzle ($He=0.0074$), which is the Helmholtz number of the experiment in~\citet{de2019generalised}. 
On the one hand, in isentropic flows, the magnitudes of $S_R$ and $S_T$ are zero throughout. 
On the other hand, when friction is modelled, both $S_R$ and $S_T$ increase in magnitude. 
In a nearly compact nozzle, there is a negligible phase difference between the reflected and transmitted waves. 
The model predictions on entropic-acoustic reflection and transmission coefficients compare favourably with the theoretical results presented in Figure 6 of \citet{de2019generalised} where $\beta$ (discussed in \S~\ref{subsec:betavsf}) was used to measure the degree of non isentropicity.
For a frictionless compact nozzle with equal inlet and exit areas, 
the acoustic-acoustic transmission coefficient is unity and the reflection coefficient is zero. 
In contrast, as the friction, $f$, increases, the reflection coefficient approaches unity for choked conditions, whereas the transmission coefficient approaches zero. 
This means that modelling friction, thereby relaxing the isentropic assumption, is key to the accurate prediction of indirect noise in subsonic nozzles.
Physically, for higher values of friction, all the impinging acoustic waves tend to reflect back. 
Figures \ref{fig:fhezero_1} (c,d) show the comparison of the model prediction on acoustic-to-acoustic transfer functions with  the experimental data of \citet{de2019generalised}. The error bars are reported in~\citet{Domenico2020}. 
A favourable fit is found for $f = 0.07-0.08$, which is a physical range of friction in ducts~\citep{shapiro1954compressible}. 

\subsection{Indirect noise in non-compact nozzles}\label{sec:irjf430ff}
\begin{figure}
\centering    
\includegraphics[width=1.0\textwidth]{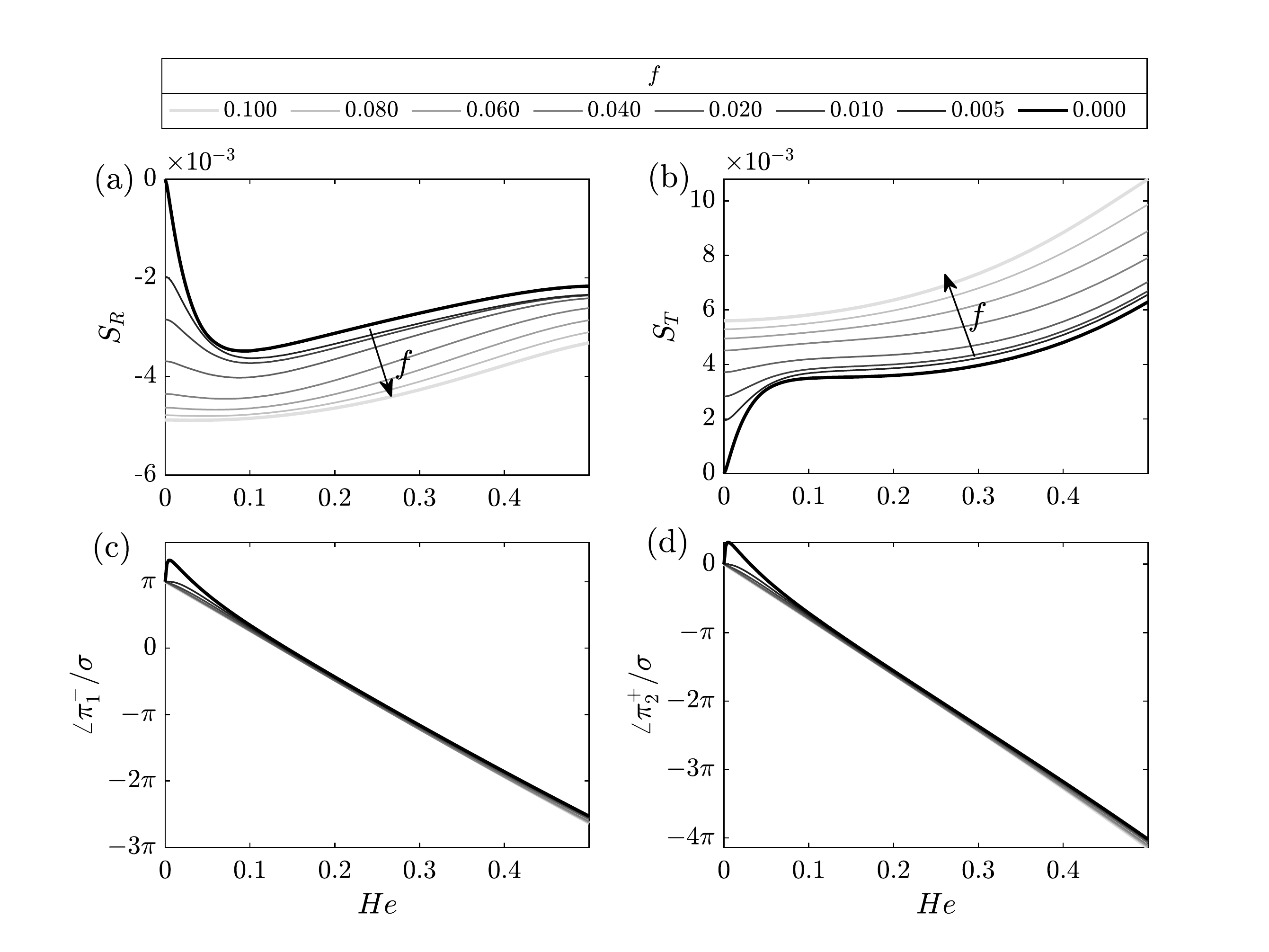}
\caption{Non-compact nozzle. Entropic-acoustic (left) reflection and (right) transmission coefficients. Gain (top) and phase (bottom). Throat Mach number $M_t = 0.8$. } 
\label{fig:fheaxis1}
\end{figure}
%
Figure \ref{fig:fheaxis1} shows the variation of the gain/phase of the indirect noise transfer functions with the Helmholtz number. 
The magnitude of the reflection coefficient, $S_R$, has a non-monotonic behaviour, which 
increases up to $He \approx 0.1$ and, then, decreases for higher $He$. 
The magnitude of the transmission coefficient, $S_T$, however, increases monotonically with $He$. 
(In the limit of a frictionless nozzle, the predictions tend to those of the isentropic model of~\citet{duran2013solution}.) 
For compact nozzles, $S_R$ and $S_T$ are zero in frictionless nozzles of equal inlet and exit areas. 
This is because the velocity gradient is equal in the convergent and divergent sections, but with different signs. 
The sound waves, thus, cancel each other. 
However, in a non-compact nozzle, the spatial extent of the velocity gradient is not negligible, which causes sound waves not to cancel each other.
Therefore, the sound is generated even in a nozzle with equal inlet and exit areas.  
As the friction increases, the amplitudes of the reflected and transmitted wave increase. 
Figure \ref{fig:fheaxis1} (c,d) show a linear change in phase as the $He$ increases. 
The peak at  small values of $He$ in frictionless nozzles is consistent with 
experimental and analytical investigation~\citep{bake2007investigation, lourier2014numerical}.
The phase of reflected waves becomes a key parameter for the prediction of thermoacoustic instabilities, as explained in \S~\ref{sec:tas}. 
 {
\section{Indirect-noise transfer functions in a supersonic regime}\label{sec:results_sup}
%
In the previous section, we show the analysis for a subsonic choked regime, which is relevant to  nozzle guide vanes of realistic aircraft engines \citep{giusti2019flow}. In this section, we extend the study to a supersonic flow with and without a normal shock. A linear steady velocity choked profile is considered in the analysis to compare the results with  \citet{duran2013solution}. The Mach numbers at the inlet and exit are fixed to $0.29$ and $1.5$, respectively. The area is calculated at each location using \eqref{eq:a2ba1m2bm1} for different friction factors. The nozzle profile for the isentropic case ($f = 0$)  is shown in Figure \ref{fig:SrsupersonicshockvsHe3}. 
\subsection{Supersonic flow without a shock wave}\label{sec:sec:results_sup_noshock}
In a choked nozzle, the mass flow rate is maximum, equivalently, the choking condition is derived by imposing zero fluctuations of Mach number at the throat \citep{marble1977acoustic}, $M^\prime/\bar M = 0$,
\begin{align}
    2\frac{u^\prime}{u} + \frac{p^\prime}{\gamma p} (1 - \gamma) - \frac{s^\prime}{c}_p = 0.
\end{align}
Friction affects the choking condition through the mean-flow quantities, but it does not affect the form of the equation. 
Because the upstream acoustic wave changes  direction at the throat, which gives rise to a singularity,  the nozzle is divided into two sections on both sides of the throat \citep{duran2013solution}. First, the flow is solved in the converging section. Second, the choking condition is imposed at the nozzle throat. Third, the flow is solved in the divergent section.
%
\begin{figure}
\centering    
\includegraphics[width=0.8\textwidth]{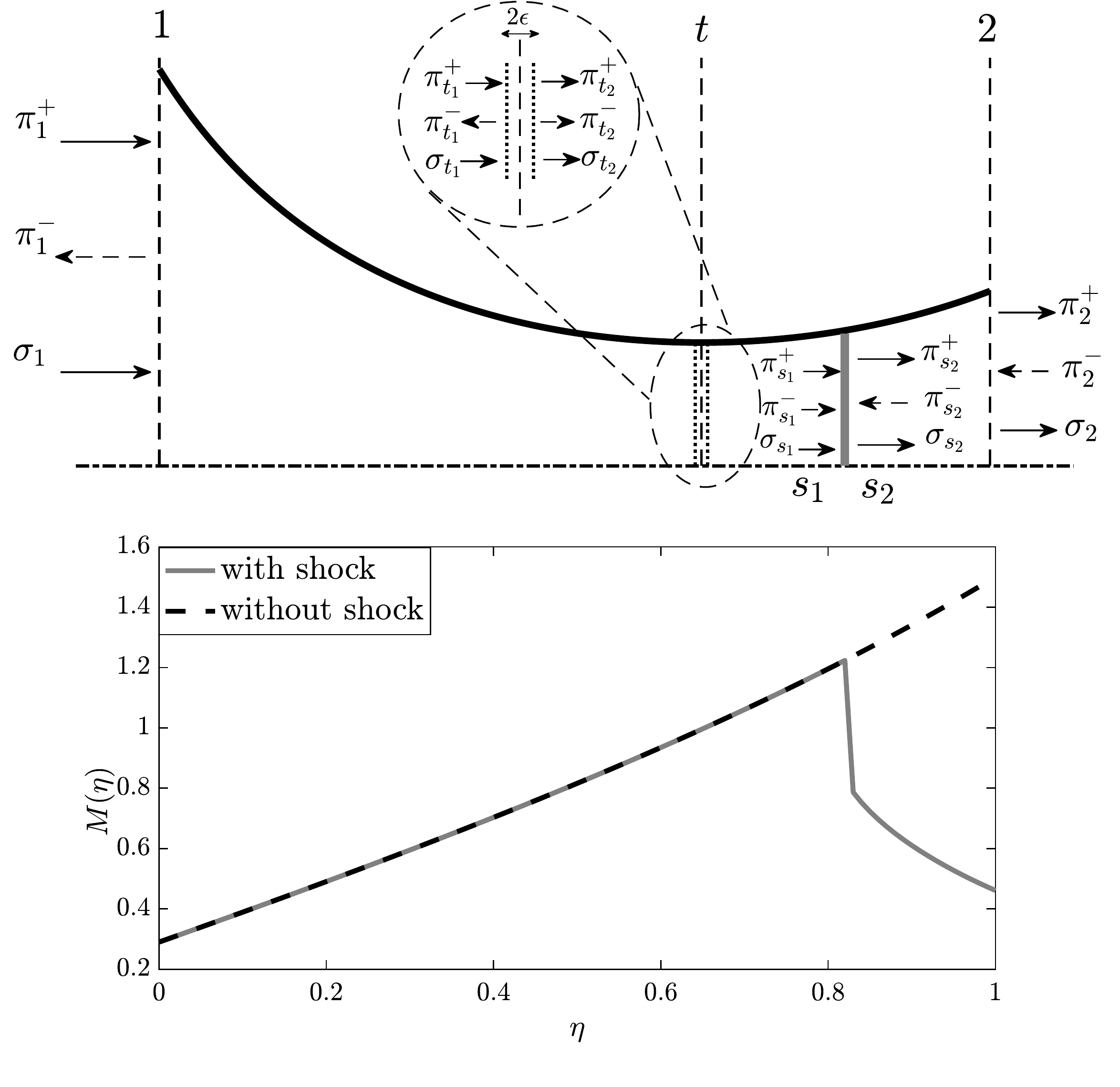}
\caption{ {(Top) Nozzle profile for the study in the supersonic regime. (Bottom) Mach number in the nozzle with steady linear velocity profile, with and without shock in the divergent section. $(M_1 = 0.29, M_2 = 1.5)$.}
}
\label{fig:SrsupersonicshockvsHe3}
\end{figure}
\begin{figure}
\centering    
\includegraphics[width=1.0\textwidth]{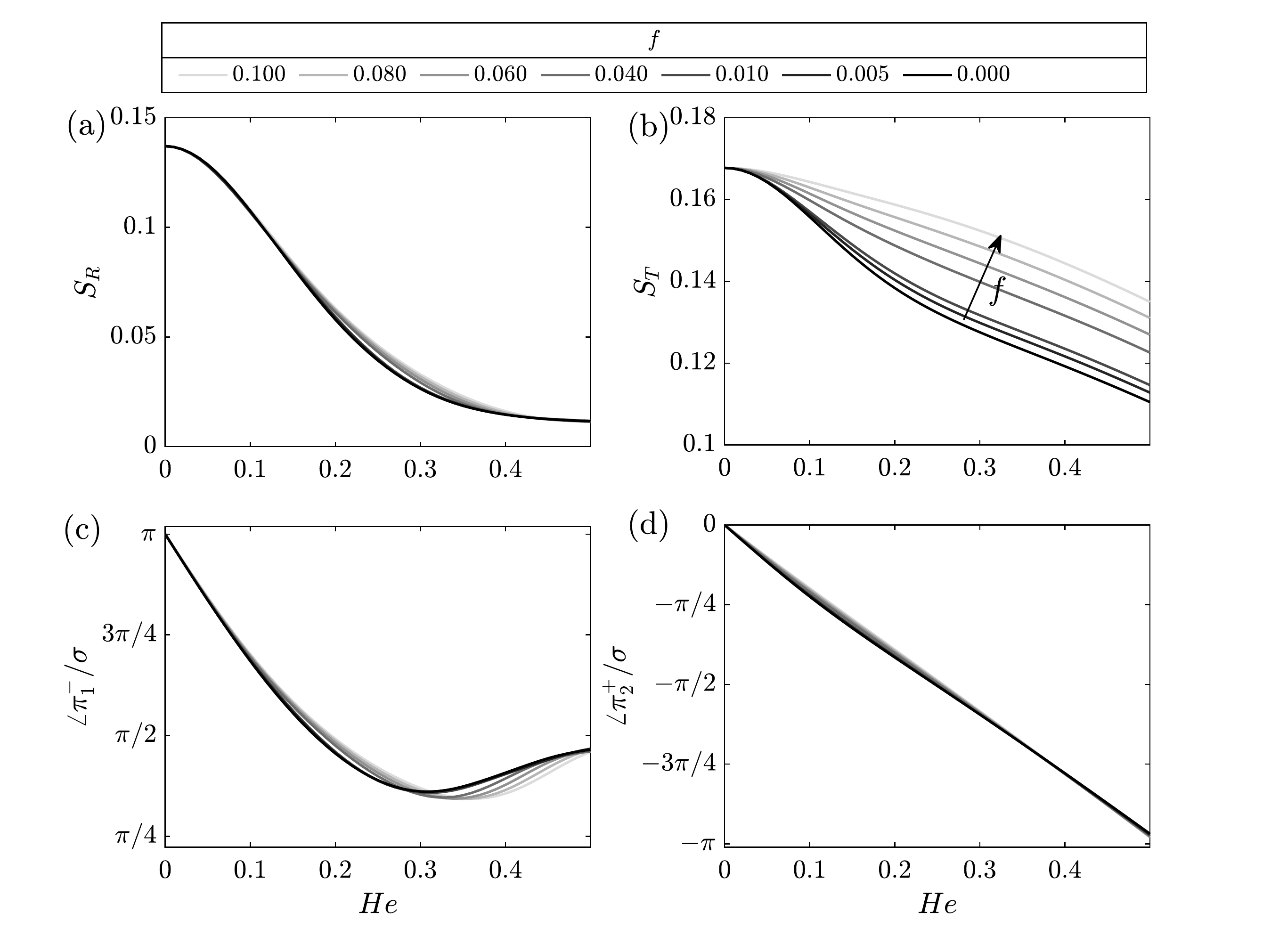}
\caption{
Non-compact nozzle. Entropic-acoustic (left) reflection and (right) transmission coefficients. Gain (top) and phase (bottom). Supersonic nozzle with a linear velocity profile without a shock wave $(M_a = 0.29, M_b = 1.5)$. 
}
\label{fig:SrsupersonicvsHe1}
\end{figure}
} 

 {Figure \ref{fig:SrsupersonicvsHe1} shows the effect that friction has on the entropic-acoustic reflected and transmitted waves. The results for the isentropic case ($f = 0$) match the results of \citet{duran2013solution} for the same nozzle profile. (The magnitudes of $S_R$ and $S_T$ are half of those of \citet{duran2013solution} because they defined the Riemann invariants without the $1/2$ factor.)
The effect of non-isentropicity is significant in the diverging section. Because the nozzle is choked, the effect of non-isentropicity in the divergent section cannot travel upstream, which means that, in contrast to the subsonic case, the reflection coefficient is almost insensitive to friction 
(Figure \ref{fig:SrsupersonicvsHe1}
(a,b)).}
 {For small Helmholtz numbers, the magnitudes of the entropic-acoustic transfer functions are insensitive to friction. As the Helmholtz number increases, friction increases the magnitude of the transmission transfer function.
In a supersonic nozzle, \eqref{eq:stagpressuremach} and table \ref{tab:stagp} show that the sign of $dp$ and $dM$ remains constant. Therefore, as expected from \eqref{eq:stagpressuremach}, the change in stagnation pressure is smaller, which  results in less entropy generation (relative to the subsonic-choked case).
}


 {In conclusion, the effect of friction in the supersonic case without a shock wave is negligible on the reflected wave for any Helmholtz number, and it is negligible on the transmission wave for small Helmholtz numbers.} 

\begin{table}
  \begin{center}
\def~{\hphantom{0}}
  \begin{tabular}{l|c|c||c|c|}
  \multicolumn{1}{c}{} &\multicolumn{2}{|c|}{Subsonic} &\multicolumn{2}{|c|}{Supersonic} \\
  \hline
        & $\eta<\eta_t$   &   $\eta>\eta_t$ & $\eta<\eta_t$   &   $\eta>\eta_t$ \\[3pt]
      $dp$   & $< 0$ & $> 0$ & $< 0$ &$< 0$\\
      $dM$   & $> 0$ & $< 0$ & $> 0$ &$> 0$\\
 \end{tabular}
  \caption{Change in pressure and Mach number for supersonic and subsonic nozzles. $\eta$ is the nozzle spatial distance, $\eta_t$ is the location of the throat.}
  \label{tab:stagp}
  \end{center}
\end{table}

 {
\subsection{Supersonic flow with a shock wave}
\begin{figure}
\centering    
\includegraphics[width=1.0\textwidth]{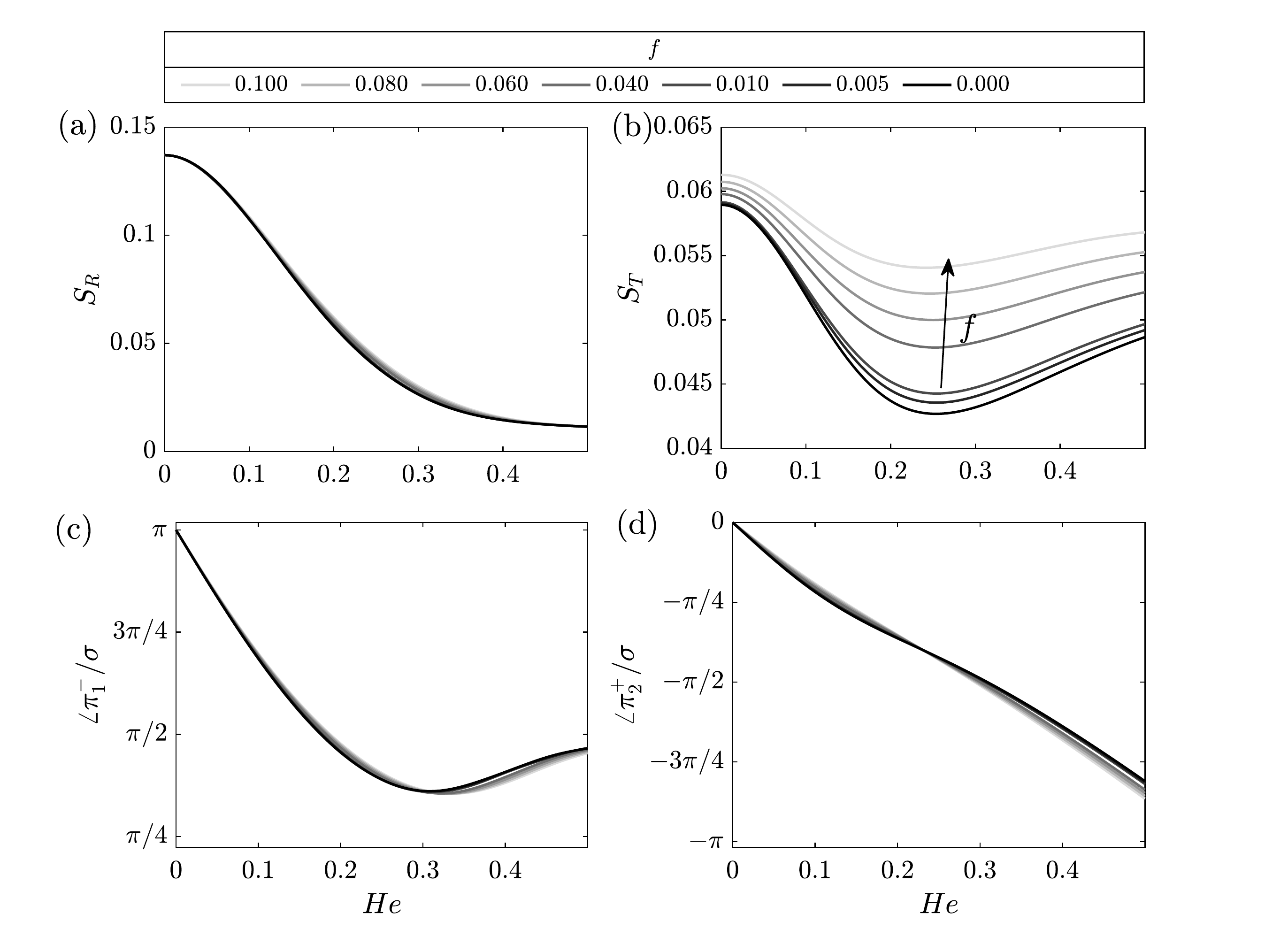}
\caption{Non-compact nozzle. Entropic-acoustic (left) reflection and (right) transmission coefficients. Gain (top) and phase (bottom). Supersonic nozzle with a linear velocity profile with a shock wave in the divergent section $(M_a = 0.29, M_b = 1.5)$. 
}
\label{fig:SrsupersonicshockvsHe1}
\end{figure}
 {
}
A normal shock wave is assumed to occur in the divergent section with  a linear velocity profile (Figure~\ref{fig:SrsupersonicshockvsHe3}). 
To solve the flow, the nozzle is divided into two parts: The flow upstream of the shock, which is calculated using the supersonic flow of \S\ref{sec:sec:results_sup_noshock}, and downstream of the shock, which is calculated with the subsonic flow conditions of \S\ref{sec:results}. The shock wave is assumed to oscillate with an infinitesimal amplitude about a mean position. Therefore, the jump conditions have the same form as the linearized Rankine-Hugoniot for frictionless nozzles, which are~\citep[e.g.,][]{marble1977acoustic, stow2002reflection, moase2007forced, goh2011phase, leyko2011numerical} 
\begin{align} \label{eq:sfhuwhrfi}
    M_{s_2}^2 &= \frac{1 + \frac{\gamma - 1}{2} M_{s_1}^2}{\gamma M_{s_1}^2 - \frac{\gamma - 1}{2}}, \nonumber \\ 
    \pi_{s_2}^+ &= \frac{1 + M_{s_2}^2 M_{s_1} + M_{s_1}^2}{1 + M_{s_1}^2 M_{s_2} + M_{s_1}^2} \pi_{s_1}^+ + \frac{1 - M_{s_2}^2 M_{s_1} + M_{s_1}^2}{1 + M_{s_1}^2 M_{s_2} + M_{s_1}^2} \pi_{s_1}^-, \nonumber \\ 
    \sigma_{s_2} &= \sigma_{s_1} + \left( \frac{(\gamma - 1)(M_{s_1} - 1)^2}{M_{s_1}^2(2 + (\gamma - 1)M_{s_1}^2)}\right)(\pi_{s_2}^+ + \pi_{s_2}^- - \pi_{s_1}^+ - \pi_{s_1}^-), 
\end{align}
where the nomenclature is defined in Figure~\ref{fig:SrsupersonicshockvsHe3}. 
Friction affects the mean flow quantities that appear in the jump conditions~\eqref{eq:sfhuwhrfi}.}
 {(In this analysis, we assume that $\pi_{s_2}^-$ is zero for brevity.) 
The nozzle response has common features with both the shockless supersonic and subsonic cases (Figure~\ref{fig:SrsupersonicshockvsHe1}). 
On the one hand, because the nozzle is choked, the reflection coefficient remains virtually unaffected by friction, as in the supersonic case without a shock (Figure~\ref{fig:SrsupersonicshockvsHe1}(a,c)). On the other hand, because the flow becomes subsonic after the shock wave, the transmission coefficient is affected by friction starting for small Helmholtz numbers, as in the subsonic case and in contrast to the shockless supersonic case (Figure~\ref{fig:SrsupersonicshockvsHe3}(b,d)). 
Friction tends to increase the magnitude of the transmitted wave, similarly to the observations in subsonic and supersonic cases~\ref{fig:SrsupersonicshockvsHe1}(c,d)). 
In the limit of zero friction, the trends of Figure~\ref{fig:SrsupersonicshockvsHe3} match qualitatively 
the results of~\citet{goh2011phase}. (We use a slightly different nozzle geometry, therefore we do not expect the results to quantitatively match.)
}


\section{Thermoacoustic stability}\label{sec:tas}


\begin{figure}
\centering    
\includegraphics[width=1.0\textwidth]{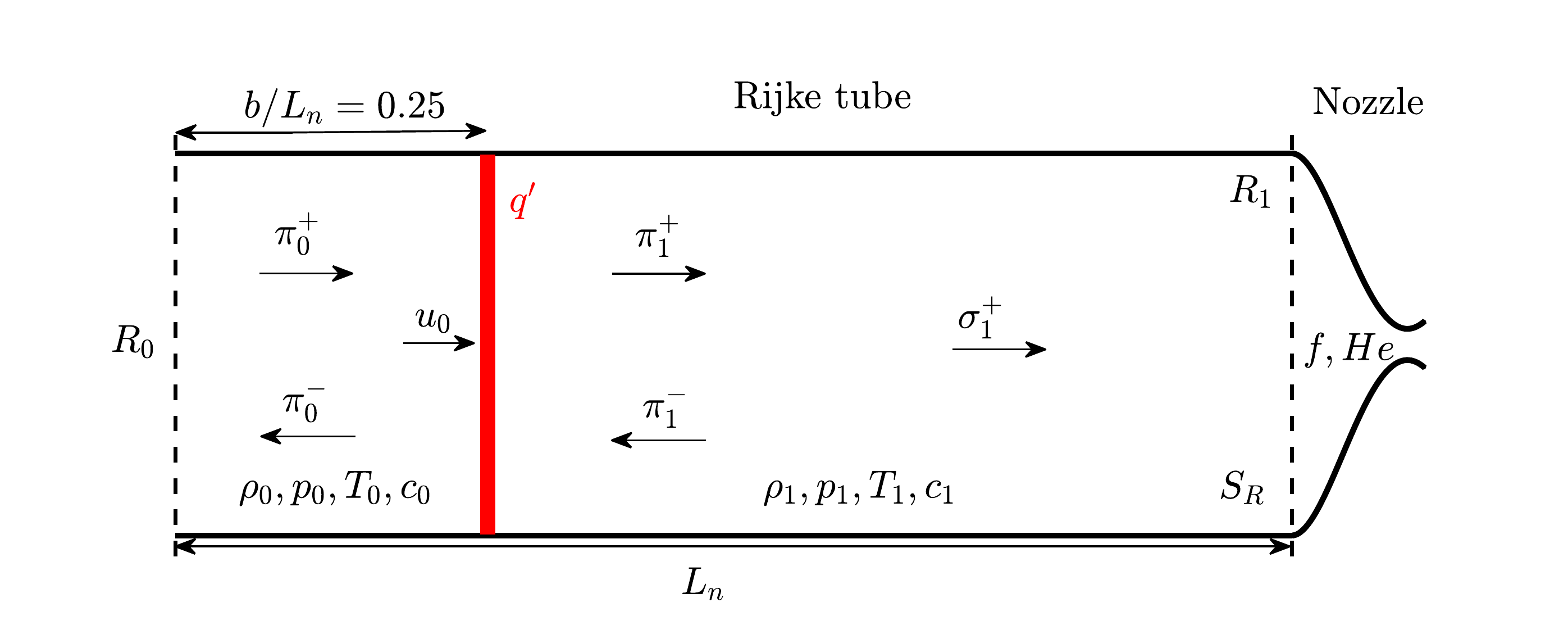}
\caption{Rijke tube with an open left end. The nozzle with dissipation is attached downstream to provide the right boundary conditions through  $S_R$, which is the acoustic-entropic transfer function, and $R_1$, which is the acoustic-acoustic transfer function.} 
\label{fig:TA_RT_1}
\end{figure}
We investigate the effect that dissipation in the nozzle guide vane has on thermoacoustic feedback. 
Geometrically, we consider a straight duct that models a laboratory combustor, also known as the Rijke tube (Figure \ref{fig:TA_RT_1}).
The duct is characterized by a left boundary condition with the acoustic-to-acoustic  reflection coefficient $R_0$, and a right boundary condition with the acoustic-to-acoustic reflection coefficient $R_1=\pi^{-}_1/\pi^+_{1}$, and entropic-to-acoustic reflection coefficient $S_R=\pi^{-}_1/\sigma_1$. 
The reflection coefficient $R_0=-1$ models an open end, which is fixed, whereas 
the reflection coefficient $R_1$ and $S_R$ are calculated from the proposed nozzle model (\S~\ref{sec:riemaninvar}), which vary with the Helmholtz number, $He$, and the nozzle friction, $f$. 
The straight duct contains a point-wise heat source, which models the heat released by a flame that responds to acoustic perturbations as 
${q^\prime} = n {u^\prime(t-\tau)}$, 
where $\tau$ is the flame time delay, 
$n$ is the flame interaction index~\citep[e.g.,][]{dowling2015combustion}. 
The conservation of mass, momentum and energy are enforced across the flame as jump conditions~\citep{bloxsidge1988reheat}. 
After Laplace transformation of the jump conditions, the thermoacoustic stability is governed by a nonlinear eigenvalue problem 
\begin{align}\label{eq:fijfr32}
\mathbf{L}(\lambda, n, \tau, M, R_0, R_1, S_R)\mathbf{q} = 0, 
\end{align}
where 
$\mathbf{L}$ is the scattering matrix, 
$\lambda$ is the complex eigenvalue, which is the solution of the dispersion relation $\det(\mathbf{L})=0$, and $\mathbf{q}$ is the eigenvector that contains the outgoing waves. 
(The exact scattering matrix $\mathbf{L}$ is reported in  Appendix B of \citet{aguilar2017adjoint}, in which all the details can be found.)
If the growth rate of the eigenvalue (real part) is positive, $\mathcal{R}(\lambda)>0$, the thermoacoustic system is linearly unstable. 
As in~\citet{aguilar2017adjoint}, we fix $n=1$ and $\tau=10^{-3}s$.
By numerically solving~\eqref{eq:fijfr32}, we compute the eigenvalue loci $\lambda(R_1(He, f), S_R(He, f))=\lambda(He, f)$ to investigate the effect that the nozzle geometry and dissipation have on thermoacoustic stability. 
The nozzle throat Mach number is fixed to $M_t=0.8$ as in $\S\ref{sec:irjf430ff}$.

%
\begin{figure}
\centering    
\includegraphics[width=1.0\textwidth]{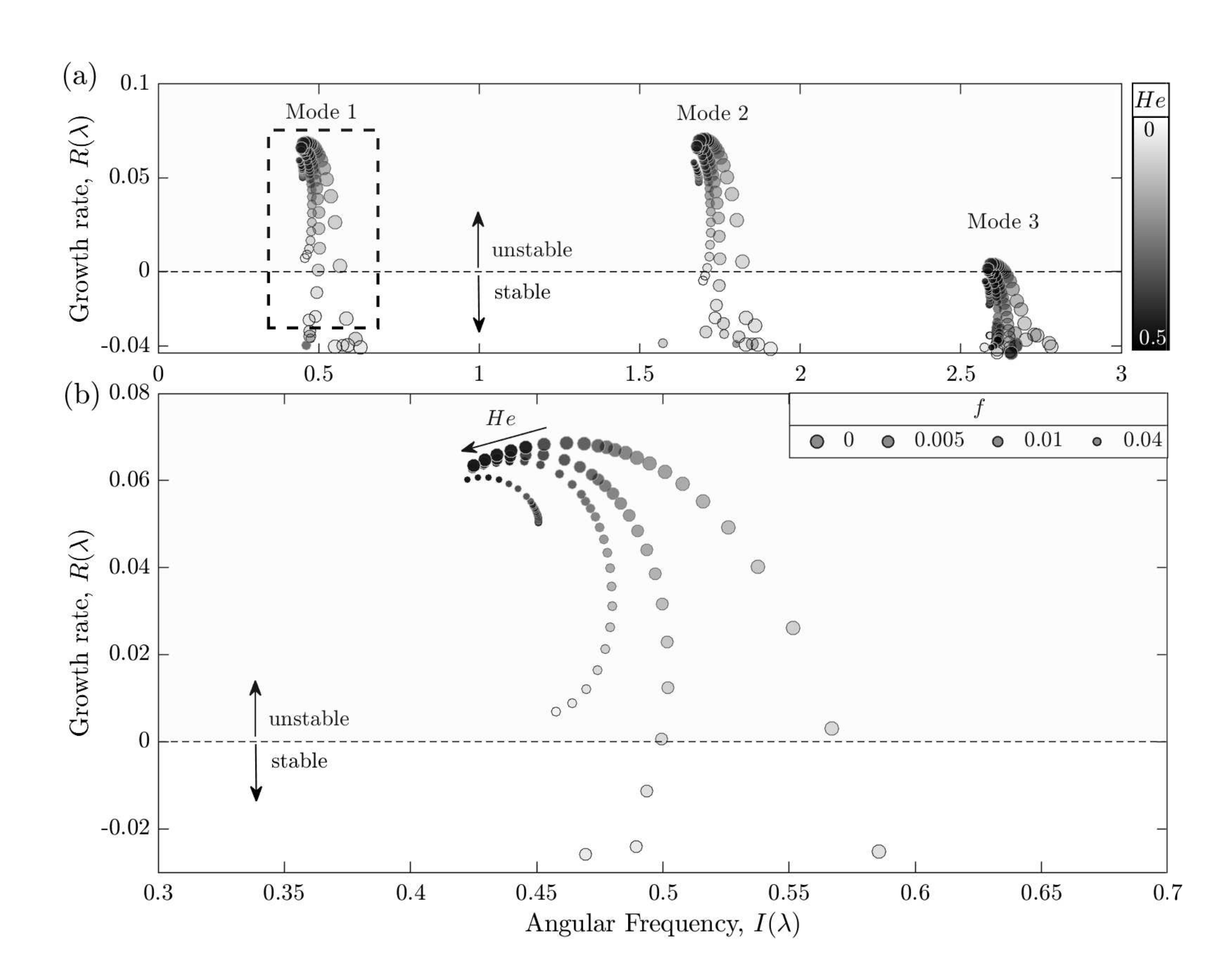}
\caption{Thermoacoustic stability (eigenvalue loci). (a) The three dominant eigenvalues (modes) for $He=0 - 0.5$ and friction factor $f = 0 - 0.04$. (b) Closeup on the first thermoacoustic mode. The axes are scaled by a factor of $2c_1/L_n=1553/1=1553s^{-1}$.} 
\label{fig:TA}
\end{figure}

Figure \ref{fig:TA} shows the trajectory of the dominant eigenvalues with respect to the nozzle guide vane friction and Helmholtz number. 
In isentropic nozzles, $f=0$, the modes become unstable as  $He$ increases. 
For a small friction, $f = 0.005$, the mode with $He = 0$ (compact nozzle) is stable. 
As the friction factor increases, all modes become unstable regardless of the Helmholtz number. 
Figure \ref{fig:TA}(b) shows that, for higher values of friction factors, the dependence of the growth rate on $He$ decreases. 
Crucially, thermoacoustic stability can switch from stable to unstable.
Physically, the nozzle Helmholtz number and friction factor change the phase of the waves that are reflected off the nozzle. 
When these waves are sufficiently in phase with the heat released by the flame, according to the Rayleigh criterion~\citep{rayleigh1896theory}, interpreted in the frequency domain~\citep{magri2020sensitivity}, the thermoacoustic system is linearly unstable. 
 {In conclusion, the analysis  shows that the friction and the spatial extent of the nozzle guide vane can have a key effect on thermoacoustic stability.}

\section{Conclusions}\label{sec:Conclusions}
Indirect noise generated in nozzles is commonly modelled with isentropic models. 
Recently, a non-isentropic model was proposed to predict the noise generated in compact nozzles~\citep{de2019generalised}.
This nozzle model is semi-empirical, i.e., it is  {based on a heuristic argument} and depends on the equivalent orifice parameter, which needs to be tuned experimentally. 
In this paper, we propose a physics-based model of indirect noise generated in subsonic and supersonic nozzles (with and without a shock wave). The model is derived from conservation laws.  
 {First, we observe that the friction factor can be used as a global variable to model the dissipation due to various factors averaged across the cross section.}
The equations for the acoustics generated by the passage of an entropy inhomogeneity are derived from first principles. 
These equations depend on the Helmholtz number, which encapsulates the spatial extent of the nozzle with respect to the acoustic wavelength. 
Second, the semi-empirical parameter of the non-isentropic model from the literature is mathematically explained with physical and measurable quantities, such as the entropy loss.
Third, we numerically solve the equations to gain physical insight into indirect noise. 
 For a fixed outlet pressure and inlet Mach number, the flow accelerates to compensate for pressure loss due to wall friction, as expected in a Fanno flow. 
The magnitude of the entropic-acoustic reflection and transmission coefficients increase with increasing levels of non-isentropicity, the effect of which becomes more significant as the throat Mach number increases.  
 The model is validated against the  experimental data available from the literature.
  {Fourth, we extend the model to supersonic flows with and without a shock wave. Friction increases the magnitude of the transmission coefficient for (i) both compact and non-compact nozzles in supersonic flows with a shock wave, and (ii) only  non-compact nozzles in supersonic flows without a shock wave.}
 {Fifth,} we show that non-isentropicity in the nozzle can have a significant effect on thermoacoustic stability. %
Systems that are thermoacoustic stable can become unstable for some friction factors and Helmholtz numbers. 
This is because the friction and the Helmholtz number have a marked effect on the phase of reflected waves, which, in turn, can fulfil the Rayleigh criterion when travelling upstream to the flame.

This work opens up new possibilities for accurate modelling of indirect noise and thermoacoustic stability in aeronautics and power generation with realistic nozzles. 
 




\section*{Acknowledgements}
{A. J. is supported by the University of Cambridge Harding Distinguished Postgraduate Scholars Programme. L.M. acknowledges the support from the Royal Academy of Engineering Research Fellowships and the ERC Starting Grant (PhyCo, no. 94938).}






\appendix

\section{Linearised entropy source term}\label{app:fihjrf329f3}

The right-hand-side term of equation \eqref{eq:dscpdt_1} is,
\begin{align}
&g(f,f^2,f^3, o(f^3)) = C_1\left(\frac{M^2(\gamma - 1)}{2}\frac{\partial}{\partial \eta}\left(\frac{u^{\prime}}{ u}\right) - \frac{\partial}{\partial \eta}\frac{p^{\prime}}{\gamma p}\right) + C_2 \left(2\frac{u^{\prime}}{ u} + (1 - \gamma) \frac{p^{\prime}}{\gamma p} - \frac{s^\prime}{c_p}\right), \\
    &C_1 = \Theta f\left(\gamma M^2 f - 2 \tan\alpha \right),\\
    &C_2 = \Theta \Bigg(2 \tan\alpha M \frac{d M}{d\eta}\left(\frac{1 - 
    (\gamma + 2) M^2}{(1 - M^2)\Lambda} + 2\right) f  \ldots \nonumber\\
    \ldots& - \left(\gamma M^3 \frac{d M}{d\eta}\left(\frac{1 - 
    (\gamma + 2) M^2}{(1 -  M^2)\Lambda}\right) - \frac{4\tan\alpha}{D} M^2 \right)f^2 - \frac{2\gamma M^4}{D}f^3\Bigg),
\end{align}
and
\begin{align}
    &\Theta = \tilde{u}\frac{2(\gamma - 1) (1 -  M^2)}{8\tan^2\alpha\Lambda - 2 \tan\alpha\gamma M^2(\gamma - 1) (1 -  M^2)f + (\gamma + 1)\gamma^2  M^4 f^2}.
\end{align}

\end{document}